\documentclass[final,5p,times,twocolumn]{elsarticle}

\usepackage{lineno,nohyperref}
\usepackage{url}
\usepackage{graphicx}
\usepackage{epstopdf}
\usepackage{dcolumn}
\usepackage{bm}
\usepackage{color}
\usepackage{epsfig}
\usepackage{amsmath}
\usepackage{amssymb}
\usepackage[caption=false]{subfig}
\modulolinenumbers[5]

\journal{arXiv}

\def\ee{\end{equation}}
\def\be{\begin{equation}}
\def\l{\left}
\def\r{\right}

\begin{document}

\begin{frontmatter}

\title{A note on the experiment parameters for the non-resonant streaming instability: competition between left and right circularly polarized modes \tnoteref{work}}

\fntext[work]{This work was first presented in the 12th International Conference on High Energy Density Laboratory Astrophysics (HEDLA18) in Kurashiki, Okayama, Japan. The respective proceedings were submitted to the Virtual Special Issue of High Energy Density Physics and were not accepted after the refereeing process.}

\author[add1]{Chun-Sung Jao}
 \ead{csjao899@gmail.com}
\author[add1,add2]{Sergei Vafin}
\author[add1]{Ye Chen}
 \ead{ye.lining.chen@desy.de}
\author[add1]{Matthias Gross}
\author[add1]{Mikhail Krasilnikov}
\author[add1]{Gregor Loisch}
\author[add3]{Timon Mehrling}
\author[add4]{Jacek Niemiec}
\author[add1]{Anne Oppelt}
\author[add5]{Alberto Martinez de la Ossa}
\author[add3]{Jens Osterhoff}
\author[add1,add2]{Martin Pohl}
 \ead{martin.pohl@desy.de}
\author[add1]{Frank Stephan}

\address[add1]{DESY, Zeuthen, Germany}
\address[add2]{Institute of Physics and Astronomy, University of Potsdam, Potsdam-Golm, Germany}
\address[add3]{DESY, Hamburg, Germany}
\address[add4]{Institute of Nuclear Physics PAN, Krakow, Poland}
\address[add5]{Institute of Experimental Physics, University Hamburg, Germany}

\begin{abstract}
A non-resonant streaming instability driven by cosmic-ray currents, also called Bell's instability, is proposed as a candidate for providing the required magnetic turbulence of efficient diffusive shock accelerations.  To demonstrate the saturation level and mechanism of the non-resonant streaming instability in a laboratory environment, we attempt to develop an experiment at the Photo Injector Test Facility at DESY, Zeuthen site (PITZ).  As an electron beam is used to replace the proton beam to carry the cosmic-ray current in our experiment, the polarization of the non-resonant streaming instability will be modified from the left-handed (LH) mode to the right-handed (RH) mode.  The theoretical instability analysis shows that the growth rate of this RH non-resonant mode may be smaller than it of the LH resonant mode. However the LH resonant mode can be ignored in our experiment while the expected wavelength is longer than the used plasma cell.   The results of PIC simulations will also support this contention and the occurrence of non-resonant streaming instability in our experiment.
\end{abstract}

\end{frontmatter}


\section{Introduction} \label{Intro}

With the progress of experimental technologies, laboratory astrophysics became an alternative method of astrophysical study besides the classical observations and numerical simulations.  Via the connection between astrophysics and laboratory experiments, we seek to expand our understanding of physical processes by simulating the extreme astrophysical environments in the laboratory \citep{Amatucci06}. 

Bell's instability is an electromagnetic streaming instability driven by cosmic-ray current, which is considered as a candidate for the amplification of the interstellar magnetic field \citep{Bell04}.  This magnetic-field amplification is required to increase the efficiency of diffusive shock acceleration by confining cosmic-ray particles in the vicinity of the shock, which allows particles to gain energy through multiple shock crossing \citep{Schure12}.  The analytical treatments based on Magnetohydrodynamics (MHD) and kinetic aspects both support the prospect of this non-resonant streaming instability in magnetic field amplification \citep{Bell04,Zirakashvili08,Amato09}. The nonlinear saturation level of the Bell's instability at a few times the strength of the homogeneous magnetic field is also found in fully kinetic particle-in-cell (PIC) simulation \citep{Niemiec10,Kobzar17}.  

To understand the problem of magnetic field amplification in astrophysics, we attempt to develop a laboratory experiment for Bell's instability at the Photo Injector Test Facility at DESY, Zeuthen site (PITZ) \citep{Jao19}.  However, numerous competing instabilities may occur in the so-called beam-plasma system \citep{Bret09}.  To understand the characteristic of those unstable modes becomes essential for the preliminary study; for instance, to distinguish between the left- and right-circularly-polarized waves (or, says, the resonant and non-resonant modes) \citep{Weidl19a, Weidl19b}.  In this paper, we will show the plan of selection for the non-resonant streaming instability in our experiment.

\section{Photo Injector Test Facility at DESY, Zeuthen site (PITZ)} \label{PITZ}

\begin{figure*}
\centering
\includegraphics[width=\textwidth]{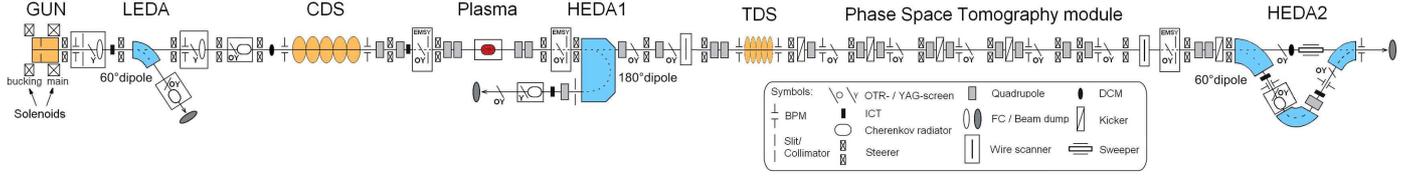}
\caption{General layout of the PITZ beam line.}
\label{Beamline}
\end{figure*}

The Photo Injector Test facility at the Zeuthen site of DESY (PITZ) was built to test, develop and experimentally optimize high-brightness photoelectron sources for superconducting-linac-driven SASE FELs, such as the Free electron LASer in Hamburg (FLASH) and the European X-ray Free Electron Laser (European XFEL)\citep{FSbook2016,ChenNIMA2018}. The main setup of PITZ consists of an L-band 1.6-Cell normal conducting (NC) RF gun, an L-band NC booster cavity, and various systems for electron-beam diagnostics, as shown in Fig. \ref{Beamline}. Reliable production of high-quality electron beams by the PITZ gun along with its multiple beam diagnostics provide excellent opportunities for laboratory astrophysics.

For the proposed laboratory experiment of Bell's instability at DESY, the PITZ gun will be used as electron source. The plasma environment for inducing the instability will be provided by the Gas Discharge Plasma (GDP) cell of PITZ \citep{GDP2017}. Based on available equipment of the PITZ accelerator, first beam-dynamics simulations have been performed for producing quasi-continuous-wave (cw) electron beams with an average beam current at the level of milliamperes and a mean beam energy of a few MeV as required for the laboratory experiments. The resulting beam parameters are: a peak beam current of 12 mA, an average current of about 2 mA, an RMS transverse beam size of about 1 mm and a mean kinetic beam energy of about 3 MeV. Such electron beams can be continuously injected into the plasma cell for a few ms, implying that the non-resonant streaming instability has to develop and saturate in a time period shorter than that. 

Here we briefly summarize the physical conditions of the beam-plasma system at PITZ  \citep{Jao19}, which will be applied in the linear instability analysis.  Our laboratory conditions allow the plasma density range from $10^{12}$ to $10^{15}$ cm$^{-3}$ within a plasma column length up to 45 cm.  The number density of the electron beam is expected to be in the range $10^{8}$ to $10^9$ cm$^{-3}$, the Lorentz factor in the range 5 to 10, and the beam radius around $1$~mm.  Since investigations of possible improvements of beam parameters are in progress \cite{Chen2018}, for the preparatory investigations,we extend the beam density range to $10^{11}$ cm$^{-3}$.  In addition, the lowest value of the external magnetic field should be about $1$~mT to avoid the effect of the geomagnetic field on the laboratory experiment, but we can be easily provide up to 0.1 T in our laboratory.

\section{Characteristics of the non-resonant streaming instabilities} \label{Theory}

\subsection{Bell's MHD treatment} \label{MHD}

The MHD treatment is first employed for the linear instability analysis.  In the Bell's model \citep{Bell04}, a cosmic-ray current is assumed to be directed along a background magnetic field, ${\bf B}_0$, into a stationary and homogeneous plasma with number density $n_0$. The current density, ${\bf j} = n_bq{\bf v}_b$, is determined by the particle density,$n_b$, the charge, $q$, and the bulk velocity of cosmic-ray particles, ${\bf v}_b$, respectively. The linear analysis indicates that a nearly purely growing, parallel propagating, non-resonant mode could be excited with peak growth rate \citep{Bell04, Zirakashvili08}
\begin{equation}
\gamma_{max} \simeq \frac{j}{2} \sqrt{\frac{\mu_0}{\rho}} \quad \quad  \text{at} \quad \quad k_{max} \simeq \frac{j}{2} \frac{\mu_0}{B_0} \quad \text{.}
\label{In1}
\end{equation}
Here $\rho = m_i n_0$ is the density of background plasma and $m_i$ is the mass of background ions.  Significant magnetic-field perturbations can be expected for $k r_{L,b} \gg 1$, where the Larmor radius of the beam particles, $r_{L,b}= \Gamma_b m_b v_b / e B_0$, is calculated using only the beam speed, $v_{b}$, and the rest mass of beam particles, $m_b$. The instability conditions can be rewritten as
\begin{equation}
\frac{\Gamma_b \rho_b v_b^2}{2} \frac{\mu_0}{B_0^2} \gg 1 \text{,} 
\label{In2}
\end{equation}
which indicates that the kinetic energy of the beam must be much larger than the initial magnetic energy.  Here $\Gamma_b=(1-v_b^2/c^2)^{-1/2}$ is the beam gamma-factor.  For example, assuming an electron beam with number density $10^9$ cm$^{-3}$ drifting into a plasma system with the beam-parallel magnetic field of $0.1$~T, a beam energy of at least $1$~GeV is required to induce Bell's instability.

The type of beam particles, electrons or ions, enters equations~(\ref{In1}) and (\ref{In2}) only through $v_{b}$ and $m_b$, and it is instructive to express these equations in terms of the electron plasma and Larmor frequency, $\omega_{pe}$ and $\omega_{ce}$, and the rest masses of plasma ions and electrons, $m_i$ and $m_e$, as well as that of the beam particles, $m_b$,
\begin{equation}
\gamma_{max} \simeq \frac{1}{2} \frac{v_b}{c} \sqrt{\frac{m_e}{m_i}} \frac{n_b}{n_e}  \omega_{pe} \quad \text{at} \quad  \lambda_{max} \simeq 4 \pi \frac{c}{v_b} \frac{n_e}{n_b} \frac{\omega_{ce}}{\omega_{pe}}   \frac{c}{\omega_{pe}} \quad \text{.} \label{GrowthRateWavenumber}
\end{equation}
As the beam-particle mass, $m_b$, does not appear in equation~(\ref{GrowthRateWavenumber}), we conclude that an electron beam can replace the proton beam, provided one considers electron and protons of the same speed, $v_b$, or with the same current density, $v_b n_b=\mathrm{const.}$. This may appear to make electron beams favorable on account of their higher speed, but
\begin{align}
 k_{max}r_{L,b} &\simeq \frac{\Gamma_b}{2} \frac{v_b^2}{c^2} \frac{m_b}{m_e} \frac{n_b}{n_e} \left(\frac{\omega_{pe}}{\omega_{ce}}\right)^2  \gg 1 \nonumber \\
 \Rightarrow & \frac{p_b v_b}{2\,m_e c^2} \frac{n_b}{n_e} \left(\frac{\omega_{pe}}{\omega_{ce}}\right)^2  \gg 1 \text{,}
\label{Criterion}
\end{align}
does contain $m_b$ explicitly, which in the second line we express using the beam momentum, $p_b$,. We see that for the same current density ($v_b n_b=\mathrm{const.}$) beam particles with higher momentum can more easily satisfy this condition. For the same beam energy that would favor proton beams.

In the MHD picture, the growth rate of the instability must be smaller than the gyro-frequency of background ions, 
\begin{equation}
\frac{\gamma_{max}}{\omega_{ci}} \simeq \frac{1}{2} \frac{v_b}{c} \sqrt{\frac{m_i}{m_e}} \frac{n_b}{n_e} \frac{\omega_{pe}}{\omega_{ce}} \ll 1 \quad \text{,}
\label{MHDBased}
\end{equation}
which is also more difficult to maintain, if an electron beam is used in the experiment, as it limits the current density carried by the beam.

\subsection{Kinetic aspect} \label{Kinetic}

To extend the linear treatment to a more general parameter regime, a kinetic theoretical model for the cold beam-plasma system is also employed for to understand the characteristics of the streaming instabilities, it also covers other instabilities in the beam-plasma system, which is helpful to understand the competition of instabilities in the nonlinear evolution processes.

Based on the electromagnetic dispersion equation, equation~(\ref{ColdDPR}) shown in Appendix A, the solutions of $\omega_{L}$, equation~(\ref{RootL}), and $\omega_R$, equation~(\ref{RootR}), can describe the left and right circular polarization wave modes, respectively.  As indicated, a variety of instabilities in the beam-plasma system with finite $v_b$ is expected in both modes.  In particular, while an electron beam is employed, the left-handed (LH) wave corresponds to the resonant instability, whereas the right-handed (RH) one to non-resonant \citep{Mikhailovskii74} (more detail in Appendix A).

\begin{figure*}
\centering
\includegraphics[width=\textwidth]{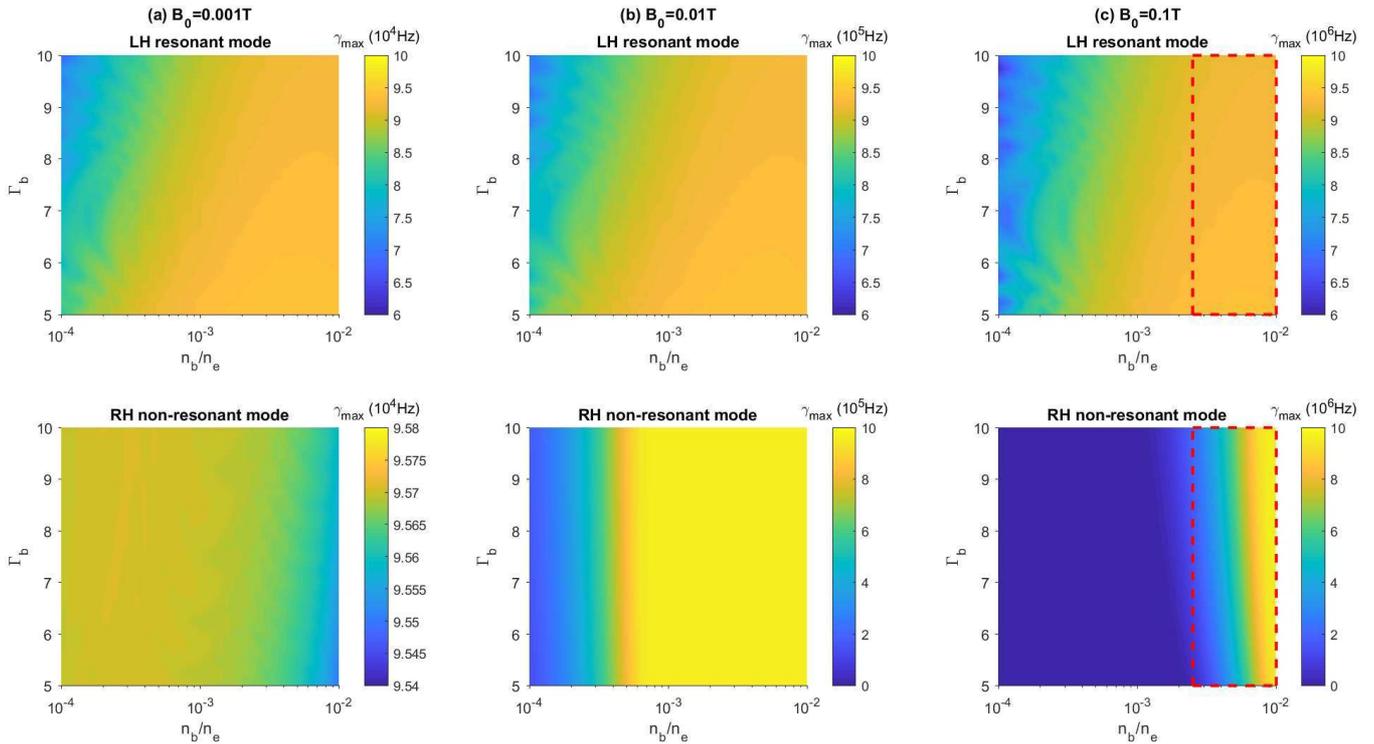}
\caption{Maximum growth rate (in units Hz) of the left-handed resonant mode (top panels) and right-handed non-resonant mode (bottom panels) of the electron beam induced electromagnetic streaming instability as function of the beam-to-plasma density ratio and beam gamma-factor with fixed parameters $n_e=10^{13}$~cm$^{-3}$ and (a) $B_0=0.001$~T , (b) $B_0=0.01$~T, and (c) $B_0=0.1$~T.  We adopt the marked parameter regime in panel (c) for the further linear analysis.}
\label{RHLH}
\end{figure*}

The maximum growth rate of the electron-beam-induced resonant streaming instability (LH mode, top panels) and non-resonant streaming instability (RH mode, bottom panels) are shown in Fig.~\ref{RHLH}.  As indicated, with increasing external magnetic field (from $B_0=$~0.001 T to 0.1 T), the linear growth rates of both the resonant and the non-resonant streaming instability increases significantly (from $\gamma_{max} \sim 10^{5}$~Hz to $\sim 10^{7}$~Hz).  As one of the most basic conditions of our experiment, the growth rate has to be large enough for the instability to develop within both the beam duration and the plasma life time. Hence a stronger external magnetic field would be beneficial for our experiment. For example for $B_0=$~0.1 T the density ratio $n_b/n_e > 2.5 \times 10^{-3}$ (the marked regime) would be required for an efficient development of the non-resonant RH mode.

\begin{figure}[!tb]
\centering
\includegraphics[width=78mm]{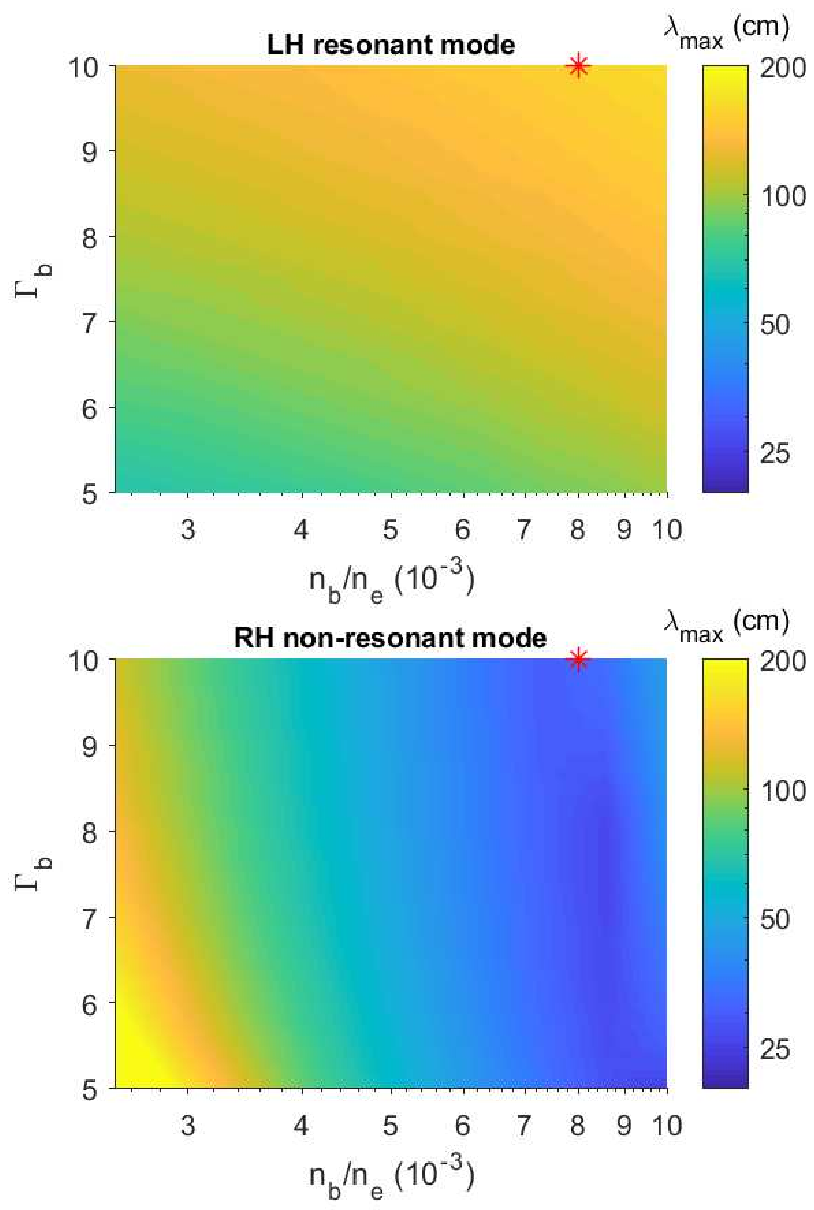}
\caption{The wavelength of the instability (in units cm) of the most unstable left-handed resonant mode (top panel) and right-handed non-resonant mode (bottom panel) of the electron beam induced electromagnetic streaming instability as function of the beam-to-plasma density ratio and beam gamma-factor.  The other (fixed) parameters are $B_0=0.1$~T and $n_e=10^{13}$~cm$^{-3}$.  We adopt the set of red-star-marked parameters for the further numerical simulation.}
\label{RHLH_L}
\end{figure}

To be noted from Fig.~\ref{RHLH}(c) is that the growth rate of the resonant LH mode can be larger than that of the non-resonant RH mode, if an electron beam is employed in our system  \cite{Amato09}.  However, as we shall see below, fitting the mode into the plasma cell imposes serious constraints on the parameters of the system.  Adopting a strong external magnetic field, $B_0=0.1$~T, and the density ratio $n_b/n_e > 2.5 \times 10^{-3}$ (the marked regime in  Fig.~\ref{RHLH}(c)), the wavelength of the most unstable resonant and non-resonant modes are shown in Fig.~\ref{RHLH_L}.  We find that the resonant LH mode perturbation has a wavelength larger than our plasma cell ($> 100$~cm~), and we will set it aside in our preliminary study.

\section{Comparison with PIC simulations}

In addition to the linear instability analysis, to understand the properties of Bell's instability induced by an electron beam, a fully relativistic, massively parallel particle-in-cell (PIC) code OSIRIS \citep{Fonseca02, Fonseca08, Fonseca13} is employed for the nonlinear calculations.  As reported in our previous study \citep{Jao19}, to mimic the laboratory environment, we assume a stationary plasma with number density $n_0 = 10^{13}$~cm$^{-3}$ within a two-dimensional periodic system in the z-x plane ($72.8$~cm~$ \times\ 3.0$~cm) and a uniform, continuous, and infinite-width cold electron beam with relativistic Maxwellian momentum distribution drifting along the $\bf{z}$ direction.  As the parameter set marked in Fig.~\ref{RHLH_L}, the constant number density and Lorentz factor of the electron beam are set to be $n_b = 8.0 \times 10^{10}$~cm$^{-3}$ and $\Gamma_b = 10$, respectively, and the background magnetic field is set as $B_0 = 0.1$~T, which is also along the $\bf{z}$ direction.  The background electrons initially drift in the $-\bf{z}$ direction for current neutrality.  To reduce the computation time, the mass ratio between ions and electrons is set to be $m_i / m_e = 50$. 

Before showing the simulation results, we perform a linear instability analysis for the simulation parameters.  Based on the MHD-based Bell's formula, the peak linear growth rate is expected to be $\gamma_{max} = 1.01 \times 10^8$~Hz at the wavelength $\lambda_{max} = 26.1$~cm.  The MHD assumption in Bell's formula is satisfied, as $\gamma_{max} / \omega_{c,i} < 1$.  With the Larmor radius of beam electrons $r_{L,b} = 17$~cm, the criterion of Bell's instability  $k_{max}r_{L,b} > 1$ is fulfilled as well. Nevertheless, the fully kinetic linear analysis reveals an unstable mode in a broad wavelength range, $\lambda > 15$~cm, with predicted peak growth rate $\gamma_{max} = 7.8 \times 10^7$~Hz at the wavelength $\lambda_{max} \simeq 26$~cm.  The predicted peak growth rate from the kinetic model is smaller than that from the MHD-based formula but the wavelengths are similar.

\begin{figure*}
\includegraphics[width=\textwidth]{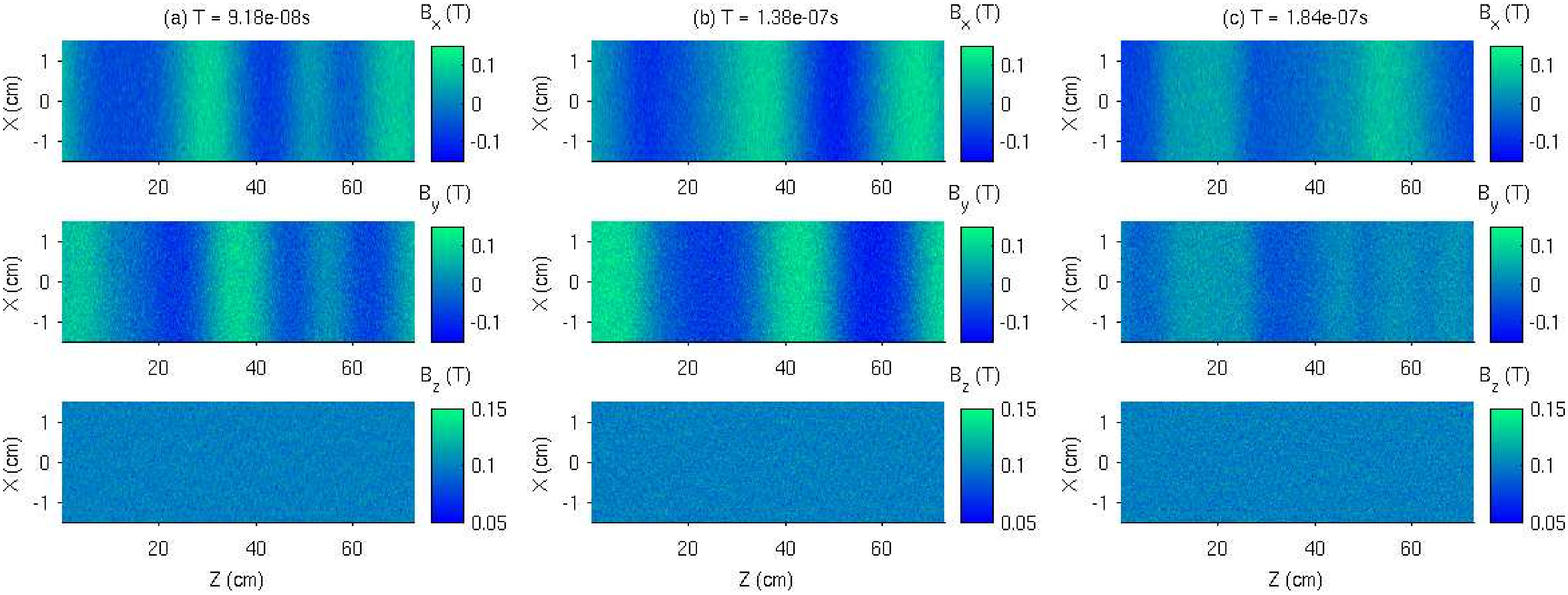}
\caption{Spatial maps of the magnetic-field components $B_x$ (top panel), $B_y$ (middle panel), and $B_z$ (bottom panel) at (a) $9.18 \times 10^{-8} s$ ($16384 \omega_{ci}^{-1}$) and (b) $1.61 \times 10^{-7} s$ ($32768 \omega_{ci}^{-1}$) for the electron-beam-induced Bell's instability.}
\label{Bell_BF}
\end{figure*}

Fig. \ref{Bell_BF} shows spatial maps of the magnetic field at $T = 9.18 \times 10^{-8} $~s ($16384 \omega_{ci}^{-1}$, panel (a)) and $1.61 \times 10^{-7} $~s ($32768 \omega_{ci}^{-1}$, panel (b)). There is no obvious structure in the field component along the homogeneous field ($B_z$).  As for the $B_x$ and $B_y$ components, the magnetic-field structures are coherent and correspond to circularly polarized plane waves travelling in the beam direction. The filamentation instability is totally suppressed by the limited size of the simulation system in the transverse direction.  Moreover, we also see that the amplitude of magnetic-field perturbations induced by Bell's instability, $B_x$ and $B_y$, can grow to the level of background magnetic field ($> 0.1 T$), but the wavelength of the dominant magnetic field perturbation changes in the nonlinear evolution processes.  By examining the time history of perturbed magnetic field at different wavelengths \citep{Jao19}, the fastest growing modes, those with wavelength $\lambda = 18.2 $~cm, $24.3 $~cm, and $36.4 $~cm, saturate around $T \sim 10^{-7} $~s, which corresponds to around $7$ (or $10$) growth times for our theoretical model (or Bell's prediction).  In particular, the wave modes with wavelength $\lambda = 36.4 $~cm has highest saturation level, and the short-wavelength mode with $\lambda = 18.2 $~cm saturates at a lower level.   As for other wave modes with more shorter wavelengths, such as $\lambda = 14.6 $~cm and $12.1 $~cm, which are stable modes in the linear analysis, the perturbations are always small and only at the level of numerical noise.

\section{Discussion and Summary} \label{Summary}

A non-resonant instability driven by cosmic-ray currents, also called Bell's instability, is a candidate for providing the magnetic-field amplification needed for efficient diffusive shock acceleration.  To understand its mechanism and properties, a laboratory experiment for Bell's instability is planned at DESY. The required electron source and plasma environment are proposed to be provided by the PITZ accelerator. The possibility of generating quasi-cw electron beams has been studied at PITZ based on a specially designed field emitter and velocity bunching by the booster cavity \citep{Chen2018}.

In our preliminary studies, we first use linear analysis to examine the most basic physical conditions that permit Bell's instability to occur in the laboratory, the linear growth rate and corresponding wavelength in particular.  Both Bell's MHD-based treatment and our kinetic theoretical model are employed, but only the kinetic model can work outside of the MHD limit and include temperature and anisotropy effects.  The kinetic model can also provide linear analysis of other instabilities in the beam-plasma system. The linear instability analysis predicts the wavelength range of the unstable modes of both left and right circular polarization modes.  

While an electron beam is employed in our beam-plasma system, the right circular polarization mode is expected to be the one corresponding to the non-resonant streaming instability.  According to our laboratory parameters, even the growth rate of the resonant LH mode is generally larger than the non-resonant RH mode, the resonant LH mode can still be set aside in our preliminary study due to its over-wavelength.  In other words, only the non-resonant RH mode can occur in our experiment while its wavelength is predicted to be far smaller than the size of our plasma cell.  The presented PIC simulations also support this contention; in particular, the simulation results shows the occurrence of the electron-beam induced Bell's instability as well as the saturation level and the spectrum of magnetic-field fluctuations, which also defend the possibility to simulate Bell's instability at PITZ.  In the context of a real experiment, the simulations are indispensable for determining the feasibility of Bell's instability in the laboratory.  By using the B-dot probe as magnetic-field detector \citep{Everson2009}, we may measure the time evolution and spatial profile of magnetic field perturbations in the plasma cell.  A comparison with the simulation data would then reveal whether or not the characteristics of the magnetic turbulence (says, growth rate and the corresponding wavelength) are consistent the properties of the fully evolved Bell's mode.

\appendix
\section{kinetic theoretical modelling for the cold beam-plasma system} \label{AppCold}

To study the electromagnetic streaming instability for parallel orientation of the wave vector with respect to the ambient homogeneous magnetic field ${\bf B}_0$ \citep{RS98,Lazar06}, the following plasma system is considered.  In the cold beam-plasma approximation, the distribution functions of the beam, background electrons and ions, respectively, are described by

\begin{equation}
f_b(p_\perp,p_z)= n_b {\delta(p_\perp)\over 2\pi p_\perp} \delta(p_z-p_b),
\label{ColdDb}
\end{equation}

\begin{equation}
f_e(p_\perp,p_z)= n_e {\delta(p_\perp)\over 2\pi p_\perp} \delta(p_z-p_e),
\label{ColdDe}
\end{equation}

\begin{equation}
f_i(p_\perp,p_z)= n_i {\delta(p_\perp)\over 2\pi p_\perp} \delta(p_z).
\label{ColdDp}
\end{equation}
The bulk momentum of the background electrons, $p_e$, drifting along the magnetic field (z-axis) is set for the current neutrality.  Then we derive the well-known dispersion equation \citep{Breizman90}

\begin{multline}
1- \l(kc\over\omega \r)^2 - {\omega_{pi}^2\over\omega(\omega\pm\omega_{ci})} -  {\omega_{pb}^2\over\omega^2}{\omega-kv_b\over\Gamma_b(\omega-kv_b\pm\omega_{cb}/\Gamma_b)} - \\
{\omega_{pe}^2\over\omega^2}{\omega-kv_e\over\Gamma_e(\omega-kv_e\pm\omega_{ce}/\Gamma_e)}=0,
\label{ColdDPR}
\end{multline}
which can be simplified to a quadratic equation
\begin{equation}
\omega^2+x_{L,R}\omega \pm x_{L,R}\omega_{ci}=0,
\label{ColdDPR_s}
\end{equation}
under the assumptions, $\l| \omega -kv_e \r|\ll \l| \omega_{ce}\r|/\Gamma_e$, $\l| kc/\omega \r|\gg1$, and $\l|\omega \r|\ll \l| kv_b\mp{\omega_{cb}/\Gamma_b} \r|$. The analytical solution of equation~(\ref{ColdDPR_s}) are 

\begin{equation}
\omega_L= {1\over2}\l[ -x_L + \sqrt{x_L(x_L+4\omega_{ci})}  \r],
\label{RootL}
\end{equation}
\begin{equation}
\omega_R= {1\over2}\l[ -x_R + \sqrt{x_R(x_R-4\omega_{ci})}  \r],
\label{RootR}
\end{equation}
where
\begin{equation} 
x_L= {\l( kV_A\r)^2\over\omega_{ci}}  + \alpha {e_b\over e_e} {\l(kv_b \r)^2\over kv_b + \omega_{cb}/\Gamma_b},
\label{X_L}
\end{equation}

\begin{equation} 
x_R=- {\l( kV_A\r)^2\over\omega_{ci}}  + \alpha {e_b\over e_e} {\l(kv_b \r)^2\over kv_b - \omega_{cb}/\Gamma_b},
\label{X_R}
\end{equation}
$\alpha=n_b/n_e$, $V_A=B_0/(4\pi n_im_i)^{1/2}$.  For $v_b=0$, equations~(\ref{RootL}) and (\ref{RootR}) describe purely oscillating Alfv\'{e}n-proton-cyclotron and whistler fluctuations, respectively.  In addition, solutions of $\omega_{L}$, equation~(\ref{RootL}), and $\omega_R$, equation~(\ref{RootR}), will be complex while $-4\omega_{ci}<x_L<0$ and $0<x_R<4\omega_{ci}$, respectively, which indicates a variety of instabilities in the beam-plasma system.

Based on equations~(\ref{RootL}) and (\ref{RootR}), we can find analytical expressions for the maximum growth rates and corresponding wavenumber (wavelength) of the unstable modes.  While the assumption $\l|\omega \r|\ll \l| kv_b\mp{\omega_{cb}/\Gamma_b} \r|$ will be easily violated for the resonant instabilities, here we will consider only non-resonant modes for the limiting case $|x_{L,R}|\lesssim 0.5\omega_{ci}$.  Thus, equations~(\ref{RootL}) and (\ref{RootR}) reduce to
\begin{equation}
\omega_L=  -{x_L\over 2} + \sqrt{\omega_{ci} x_L}  ,
\label{RootL_s}
\end{equation}
\begin{equation}
\omega_R=  -{x_R\over2} + \sqrt{-\omega_{ci} x_R}  .
\label{RootR_s}
\end{equation}
Then the instability conditions are simply $x_L<0$ and $x_R>0$, which can be evaluated as the instability conditions for an electron (RH mode) and proton (LH mode) beam, respectively,
\begin{equation}
0<k<{\omega_{ce}\over\Gamma_bv_b}+\alpha {\omega_{ci}\over v_b}\l(v_b\over V_A\r)^2, and
\label{KRange_R_E}
\end{equation}

\begin{equation}
0<k<-{\omega_{ci}\over\Gamma_bv_b}+\alpha {\omega_{ci}\over v_b}\l(v_b\over V_A\r)^2.
\label{KRange_L_P}
\end{equation}
Equations~(\ref{KRange_R_E}) and (\ref{KRange_L_P}) can both be inferred as the instability conditions shown in equations (\ref{In2}).  The maximum growth rates are
\begin{equation}
\gamma_{max,L,R} = k_{L,R} V_A \l[ 3-\sqrt{1+2/a_{L,R}} \over 1+\sqrt{1+2/a_{L,R}} \r]^{1/2},
\label{MaxGro}
\end{equation}
where
\begin{equation}
k_L= {\omega_{ci}\over v_b\Gamma_b} \l[-1 + a_L \l( 1+\sqrt{1+2/a_L} \r)  \r] ,
\label{K_L_P}
\end{equation}
\begin{equation}
k_R= {|\omega_{ce}|\over v_b\Gamma_b} \l[-1 + a_R \l( 1+\sqrt{1+2/a_R} \r)  \r] ,
\label{K_R_E}
\end{equation}
\begin{equation}
a_L= {\alpha\Gamma_b m_{p}\over 4m_i}\l(v_b\over V_A \r)^2, 
\label{A_L_P}
\end{equation}
\begin{equation}
a_R= {\alpha\Gamma_b m_{e}\over 4m_i}\l(v_b\over V_A \r)^2.
\label{A_R_E}
\end{equation}
Additionally, we have assumed that the background plasma is singly ionized which is the case in many applications. For $a_{L,R}\gg1$, we find
\begin{equation}
\gamma_{max,L,R} = k_{L,R} V_A ,
\label{MaxGro_s}
\end{equation}
\begin{equation}
k_L= {2a_L\over \rho_p}  ,
\label{K_L_P_s}
\end{equation}
\begin{equation}
k_R= {2a_R\over\rho_e} ,
\label{K_R_E_s}
\end{equation}
in agreement with equations~(\ref{In1}) and (\ref{GrowthRateWavenumber}) in the Bell's MHD treatment.

\section*{Acknowledgements}
This research is supported by the Deutsches Elektronen-Synchrotron (DESY) Strategy Fund.  The computational resources for the numerical simulations are provided by DESY and the Norddeutsche Verbund zur F\"orderung des Hoch- und H\"ochstleistungsrechnens (HLRN).  The work of J. N. is supported by Narodowe Centrum Nauki through research project DEC-2013/10/E/ST9/00662.

\bibliography{main}

\end{document}